\documentclass[aps]{revtex4} 
\usepackage{mathrsfs}
\usepackage{bm}
\usepackage{amsmath}

\begin{document}
\title{Magnetic dipole moments in single and coupled split-ring resonators}
\author{Yong Zeng, Colm Dineen and Jerome V. Moloney}
\affiliation{Arizona Center for Mathematical Sciences, University of
Arizona, Tucson, Arizona 85721}
\input epsf
\begin{abstract}
We examine the role of magnetic dipoles in single and coupled pairs
of metallic split-ring resonators by numerically computing their
magnitude and examining their relative contributions to the
scattering cross section. We demonstrate that magnetic dipoles can
strongly influence the scattering cross section along particular
directions. It is also found that the magnetic dipole parallel to
the incident magnetic field and/or high-order multipoles may play a
significant role in the linear response of coupled split-ring
resonators.
\end{abstract}
\maketitle

\section{Introduction}

In a seminal paper published in 1999, Pendry \textit{et al.}
proposed that an array of metallic split-ring resonators (SRRs) can,
in the long-wavelength approximation, possess an effective negative
permeability at its inductor-capacitor resonance \cite{john1}. This
proposal lead to the development of an artificial structured medium,
a metamaterial, consisting of interspaced SRR and metallic rods
which simultaneously displays negative permeability and permittivity
\cite{smith}. Such novel metamaterials are absent in nature and have
potential application such as subwavelength or even perfect imaging
\cite{john2,marques,solymar}. Metallic SRR arrays are a popular
building block for metamaterials and have attracted intensive
attention in the last decade \cite{marques,solymar}. Such arrays are
sometimes referred to as magnetic metamaterials due to the
significant role played by magnetic dipoles in their optical
response.

Experimentally single- and multi- layered metallic single-slit SRR
membranes have been fabricated and their optical responses are
usually studied by measuring their transmission spectra at normal
incidence \cite{linden,liu,gansel}. A magnetic dipole moment
perpendicular to the membrane, arising from a current circulating
inside the SRR, is frequently mentioned as having a role in the
observed transmission spectrum and in the coupling between adjacent
SRRs \cite{liu,liu2,schneider,hesmer}. However we propose that the
role of this magnetic dipole is unclear or even over emphasized in
the literature perhaps due to the difficulty in experimentally measuring
magnetic dipoles. Furthermore, rarely mentioned is the contribution
from a magnetic dipole parallel to the incident magnetic field
arising from the strongly localized conduction electrons near the
metallic surfaces \cite{jackson}.

The main aim of this article is to clarify the role of these
magnetic dipoles by numerically computing their magnitude and
examining their relative contributions to the scattering cross
section. To achieve our purposes, we simulate the optical response
of an isolated single SRR and a coupled pair of SRRs using the
finite-difference time-domain (FDTD) method \cite{taflove}. Our
numerical results show that the strength of the magnetic dipoles can
be determined by measuring the scattering cross section along
particular directions. We also note that in certain circumstances
the magnetic dipole parallel to the membrane plays a significant
role. For isolated pairs of coupled SRRs high-order multipoles
contribute considerately to the scattering cross section. Finally we
numerically verify the qualitative assumption in earlier literature
\cite{liu} that the magnetic dipole-dipole interaction must be
considered to interpret the coupling strength between coupled SRRs.

The papers is arranged as follows. In Section II we show that in the
presence of an electric dipole the analytical expression for a
magnetic dipole is not translational invariant. In section III we
numerically compute the various electric and magnetic dipole moments
and present a quantitative analysis of their relative contributions
to the scattering cross section. The role of magnetic dipoles and
higher order multipoles in second harmonic generation from an
individual SRR are studied in an appendix.

\section{Multipole theory in electromagnetic scattering and radiation}

Let us first recall basic multipole theory in electromagnetic
scattering and radiation \cite{jackson}. For a localized current
source $\mathbf{J}$, with sinusodial time dependence, the relevant
vector potential at position $\mathbf{r}$ is given by
\begin{equation}
\mathbf{A}(\mathbf{r})=\frac{\mu_{0}}{4\pi}\int_{v}\mathbf{J}(\mathbf{r'})\frac{e^{ik|\mathbf{r}-\mathbf{r'}|}}{|\mathbf{r}-\mathbf{r'}|}d\mathbf{r'},
\label{eq1}
\end{equation}
where $k=\omega/c$ is the wave number, and the integration is
performed over the current source. Its asymptotic form
\begin{equation}
\lim_{r\rightarrow\infty}\mathbf{A}(\mathbf{r})\approx\frac{\mu_{0}}{4\pi
r}e^{ikr}\int_{v}\mathbf{J}(\mathbf{r'})e^{-ik\mathbf{n}\cdot\mathbf{r'}}d\mathbf{r'}\equiv\frac{\mu_{0}}{4\pi
r}e^{ikr}\mathbf{p} \label{eq2}
\end{equation}
can be achieved by approximating $|\mathbf{r}-\mathbf{r'}|$ as $r$
in the denominator while in the numerator $|\mathbf{r}-\mathbf{r'}|$
is replaced by $r-\mathbf{n}\cdot\mathbf{r'}$ with
$\mathbf{n}=\mathbf{r}/r$ being the observation direction.
Consequently, the far-zone electromagnetic fields can be formulated
as
\begin{equation}
\mathbf{B}_{s}(\mathbf{r})=\frac{ik\mu_{0}}{4\pi
r}e^{ikr}\mathbf{n}\times\mathbf{p},\:\:\:\:\:\:
\mathbf{E}_{s}(\mathbf{r})= c\mathbf{B}_{s}\times\mathbf{n}.
\label{eq3}
\end{equation}
We can now write the time-averaged power scattered per unit solid
angle as \cite{yong1}
\begin{equation}
\frac{dP}{d\Omega}=\frac{1}{2}\textrm{Re}\left[r^{2}\mathbf{n}\cdot\mathbf{E}_{s}\times\mathbf{H}_{s}^{\ast}\right]=\frac{k^{2}\eta_{0}
}{32\pi^2}\left|(\mathbf{n}\times\mathbf{p})\times\mathbf{n}\right|^{2}
\label{eq4}
\end{equation}
with $\eta_{0}=\sqrt{\mu_{0}/\epsilon_{0}}$ being the intrinsic
impedance of free space. In addition, according to the optical
theorem the extinction power $P_{ex}$ taken from an incident wave
$\mathbf{E}_{0}e^{i\mathbf{k}_{0}\cdot\mathbf{r}}$ takes the form
\cite{yang}
\begin{equation}
P_{ex} =
\frac{1}{2}\textrm{Re}\left[\mathbf{E}_{0}^{\ast}\cdot\int_{v}\mathbf{J}(\mathbf{r}')e^{-i\mathbf{k}_{0}\cdot\mathbf{r}'}d\mathbf{r}'\right]=\frac{1}{2}\textrm{Re}\left[\mathbf{E}_{0}^{\ast}\cdot\mathbf{p}(\mathbf{k}_{0})\right].
\label{eq5}
\end{equation}
This expression suggests that $P_{ex}$ can be interpreted as
interference between the incident wave and the forward scattering
wave \cite{mishchenko}.

The derivations above clearly demonstrate that the
direction-dependent vector $\mathbf{p}$ contains all the information
for scattering and/or radiation. Using Taylor expansion, we can
reformulate $\mathbf{p}$ as
\begin{equation}
\mathbf{p}=\sum_{n}\frac{(-ik)^{n}}{n!}\int_{v}\mathbf{j}(\mathbf{r'})\left(\mathbf{n}\cdot\mathbf{r'}\right)^{n}d\mathbf{r'}.
\label{eq6}
\end{equation}
Mathematically this transformation is always valid, however, to
achieve physically meaningful quantities the spatial dimensions of the
current volume must be much smaller than the incident wavelength. In
other words, the quantity $k\left(\mathbf{n}\cdot\mathbf{r'}\right)$
should be small compared to unity, and consequently the successive
terms in the expansion fall off rapidly with $n$. The first term
represents the electric dipole moment,
\begin{equation}
\mathbf{d}=\int_{v}\mathbf{j}(\mathbf{r'})d\mathbf{r'}, \label{eq7}
\end{equation}
which corresponds to the electrostatic limit. The antisymmetric part
of the second term is given by
\begin{equation}
ik\mathbf{n}\times\frac{1}{2}\int_{v}\mathbf{r}'\times\mathbf{j}(\mathbf{r}')d\mathbf{r}'\equiv
ik\mathbf{n}\times\mathbf{m},\label{eq8}
\end{equation}
with $\mathbf{m}$ standing for the magnetic dipole moment, while the
symmetric counterpart corresponds to electric quadrupole
\cite{jackson}.

Despite the fact that these definitions can be found from any
popular undergraduate electromagnetism textbook, determining the
electric and magnetic dipole of a realistic system is not
straightforward. The main reason is that as long as $\mathbf{d}$
exists $\mathbf{m}$ is not translational invariant (independent of
the choice of origin of coordinates). More specifically rewriting
the above in a new coordinate system $\mathbf{x}$, which connects to
the original as $\mathbf{x}=\mathbf{r}+\mathbf{a}$, we obtain an
identical far-zone vector potential $\mathbf{A}$ but a slightly
different $\mathbf{p}$ given by
\begin{equation}
\mathbf{p}=\int_{v}\mathbf{J}(\mathbf{r'})e^{-ik\mathbf{n}\cdot(\mathbf{r'}+\mathbf{a})}d\mathbf{r'}.
\label{eq9}
\end{equation}
In this new coordinate system the electric dipole $\mathbf{d}$ is
identical to that defined in Equation \ref{eq7}, however, the
magnetic dipole $\mathbf{m}$ must be reformulated as
\begin{equation}
\mathbf{m}=
\frac{1}{2}\int_{v}\mathbf{r}'\times\mathbf{J}(\mathbf{r}')d\mathbf{r}'+\frac{1}{2}\mathbf{a}\times\mathbf{d}.
\label{eq10}
\end{equation}
The new formulation of the magnetic dipole $\mathbf{m}$ depends on
$\mathbf{a}$ and is therefore not translationaly invariant, similarly
for the high-order multipoles \cite{raab,papasimakis}. In order to
compare the relative contributions of the magnetic dipoles to the
linear scattering we therefore take the advantage of the structural
symmetries by choosing the \textit{center of mass} as the coordinate
origin.

We apply this analysis to the configurations of gold SRRs embedded
in vacuum as shown in Figure 1. The structures are excited by a
$y$-polarized plane wave propagating along the $x$ direction. It
excites not only a $y$-component electric dipole $d_{y}$, but also a
magnetic dipole possessing both $z$ and $x$ components. The $z$
component $m_{z}$ is due to the strongly localized conduction
electrons of the metal, while $m_{x}$ is induced by the structural
symmetry breaking. In order to simplify our analysis we ignore
higher order multipoles and only consider contributions from the
three dipoles $d_{y}$, $m_{x}$ and $m_{z}$.

From Equation \ref{eq4} the scattering cross section along the three
primary axes are given by
\begin{equation}
\begin{cases}|d_{y}\mp ikm_{z}|^{2}, &\text{if $\mathbf{n}$ along the $\pm x$ direction;}\\
k^{2}\left(|m_{x}|^{2}+|m_{z}|^{2}\right),&\text{if $\mathbf{n}$ along the $\pm y$ direction;}\\
|d_{y}\pm ikm_{x}|^{2}, &\text{if $\mathbf{n}$ along the $\pm z$ direction}\\
\end{cases}
\label{eq11}
\end{equation}
For example, along the $\pm x$ direction the forward- or backward-
scattering strength is influenced by the interference between the
$d_{y}$ and $m_{z}$ dipoles, while the $x$-component magnetic dipole
$m_{x}$ does not contribute. This observation is at odds with the
commonly stated assumption that the $m_{x}$ dipole plays a
significant role in one valley of the transmission spectrum of SRR
arrays where it is often referred to as the magnetic resonance
\cite{linden}. Similarly, along the $y$ direction the scattering
cross section is due solely to the $m_{x}$ and $m_{z}$ dipoles and
receives no contribution from the electric dipole $d_{y}$. This
observation may provide an insight on how to determine the relative
magnitude of the various magnetic dipoles by experimentally
measuring the scattering cross section along the orthogonal axes.

\begin{figure}
\epsfxsize=300pt \epsfbox{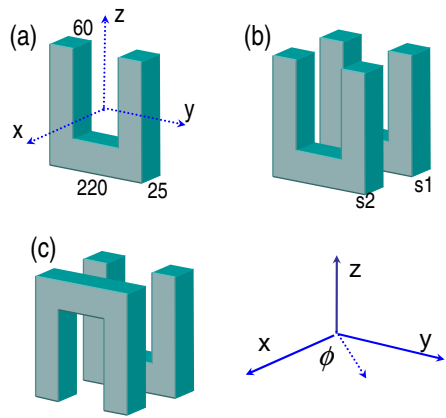} \vspace*{-8.0cm}
\caption{Schematic of (a) an individual split-ring resonator and two
coupled split-ring resonators arranged in (b) Parallel configuration
and (c) Opposite configuration. The separation between $s1$ and $s2$
is 30 nm. All other dimensions shown are in nanometers.}
\label{fig1}
\end{figure}

\section{Results and discussions}
To quantify the relative contributions of the various electric and
magnetic dipole moments we perform a numerical simulation of the
structures shown in Figure \ref{fig1} using the FDTD method. The
relative dielectric constant of gold is fitted by the Drude model of
\begin{equation}
\epsilon(\omega)=1.0-\frac{\omega_{p}^{2}}{\omega(\omega+i\gamma)},
\label{eq12}
\end{equation}
where $\omega_{p}=1.367\times 10^{16}$ s$^{-1}$ is the bulk plasma
frequency determined by the density of conduction electrons,
$\gamma=6.478\times 10^{13}$ s$^{-1}$ represents the
phenomenological damping rate \cite{palik}. Using FDTD, with a
center of mass coordinate system, we compute the polarization
current inside the metallic structures and employ Equations
\ref{eq7} and \ref{eq8} to determine the magnitude of the electric
and magnetic dipole moments. Using Equations \ref{eq4} and
\ref{eq5}, which are independent of the chosen coordinate system, we
compute the scattering and extinction cross sections \cite{yong1}.

As before the structures are excited with a plane wave propagating
along the positive $x$ direction and polarized in the $y$ direction.
The absolute extinction cross section $\sigma_{\rm{ext}}$ of the
various SRR structures are shown in Figure (2a). For the single SRR
(solid line) a peak corresponding to the fundamental plasmonic
resonance was found at a wavelength of 1264 nm. A similar individual
SRR is experimentally studied in Ref.\cite{husnik} and the
experimental extinction cross section and our numerical results are
in excellent agreement. A periodic array of SRRs with lattice
spacing of 305 nm was shown to have a fundamental resonance at 1212
nm wavelength, which is very close to the isolated SSR resonance
indicating a weak coupling between adjacent SRRs \cite{yong3}.

The scattering cross section $\sigma_{\rm{sc}}$ of the isolated SRR,
at 1264 nm wavelength, as a function of the azimuthal angle $\phi$
in the $xy$ plane is shown in Figure (2b) (solid line). This
dependence curve can be fitted by $\cos^{2}\phi$ which suggests that
the linear scattering of this individual SRR is dominated by a
$y$-component electric dipole moment $d_{y}$. Based on the analysis
related to Equation \ref{eq11} we can further estimate the magnitude
of the magnetic dipoles from the scattering along the $x$ and $y$
directions. The plotted values of $\sigma_{\rm{sc}}$ along $\pm x$
directions $[\theta = 0,180]$ are almost identical indicting a
negligible $z$-component magnetic dipole $m_{z}$. Similarly for
scattering along the $\pm y$ directions $[\theta = 90,270]$ the
cross section is nonzero indicating the presence of a finite total
magnetic dipole. Consequently we can deduce that the SRR possesses a
considerable $x$-component magnetic dipole $m_{x}$.

\begin{figure}
\epsfxsize=300pt \epsfbox{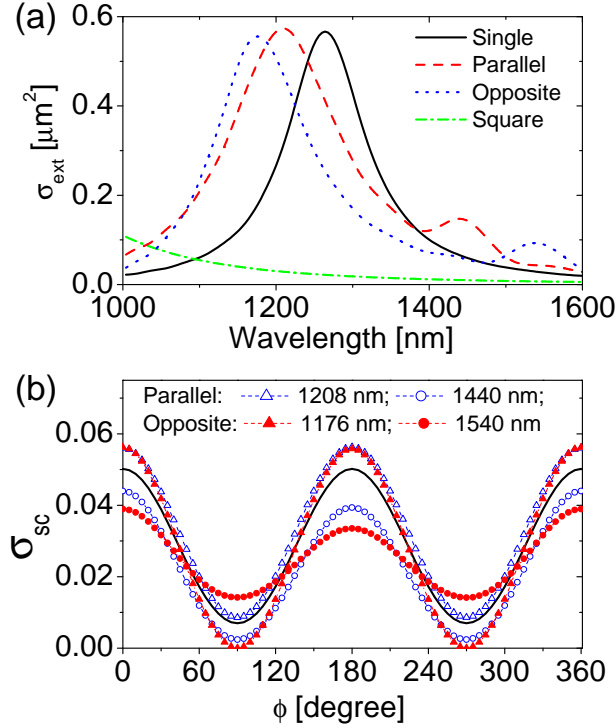} \vspace*{-3.0cm} \caption{(a)
Absolute extinction cross sections $\sigma_{\rm{ext}}$. (b)
Differential scattering cross section $\sigma_{\rm{sc}}$, with
dimensions of $\mu m^{2}$ per unit solid angle, as a function of the
azimuthal angle $\phi$. For the sake of clarity, the curves of 1440
nm wavelength and 1540 nm wavelength are amplified 10 times. The
solid black curve corresponds to the individual split-ring resonator
at 1264 nm wavelength. The incident plane wave propagates along the
positive $x$ $[\phi=0]$ direction and is $y$-polarized $[\phi=90]$.}
\label{fig2}
\end{figure}

To compare the relative strength of the numerically computed
electric and magnetic dipole moments we normalize them to the
electric dipole moment $d_{s}$ of the single SRR as follows
\begin{equation} \alpha=\left|d_{y}/d_{s}\right|,
\:\:\:\:\beta_{x}=k\left|m_{x}/d_{s}\right|,\:\:\:\:\beta_{z}=k\left|m_{z}/d_{s}\right|,
\end{equation}
and the results are shown in Table \ref{table1}. It is found that,
for the individual SRR, $\beta_{x}=0.25$ while $\beta_{z}$ is almost
zero, in agreement with the scattering-cross-section-based analysis
above. For comparison we also plot in Figure (2a) the extinction
cross section of a gold square of size $220\times220$ nm. In the
wavelength range of 1000 nm to 1600 nm the square exhibits no
resonance peaks and has a weak extinction cross section consistent
with its small normalized electric dipole of 0.23. However, it
exhibits a larger $z$-component magnetic dipole $m_{z}$ than the
single SRR and, because of its symmetry, does not present a
$x$-component magnetic dipole $m_{x}$.

We proceed to coupled SRRs oriented as depicted in Figure (1b,1c)
with a separation between $s1$ and $s2$ of 30 nm. Two different
configurations are studied. The two SRRs are parallel to each
other in the first arrangement (named Parallel) while in the
second arrangement s2 is rotated 180 degree in the $yz$ plane (named
Opposite). Their extinction spectra are shown in Figure (2a). For
both configurations the 1264 nm peak of the individual SRR is split
into two peaks induced by inter-SRR coupling. For the
Parallel structure the peaks are located at 1208 nm and 1440 nm, and
for the Opposite structure they are lactated at 1176 nm and 1540 nm.
The larger separation between the peaks of the Opposite
configuration indicates stronger inter-SRR coupling \cite{sakurai}.
Furthermore, for each structure the longer wavelength peak has a
magnitude comparable to that of the individual SRR, while the
shorter wavelength peak has significantly reduced extinction cross
sections.

For each configuration the extinction cross section as a function of
the angle $\phi$ in the $xy$ plane is shown in Figure (2b). The
scattering cross section along the $y$ direction indicates the
strength of the total magnetic dipole moment. Along the $y$
direction $[\phi = 90,270]$ the Parallel configuration at 1208 nm
possesses a scattering cross section comparable to that of the
single SRR. This is confirmed in Table \ref{table1} where
$\beta_{x}$ is 0.25 and 0.26 for the single SRR and Parallel
configuration, respectively. For the Opposite configuration at 1176
nm, however, scattering along the $y$ direction is almost zero
indicating a negligible total magnetic dipole. This is also
confirmed in the numerically computed values where $\beta_{z}$ is
0.008 and $\beta_{x}$ is an order of magnitude smaller than the
single SRR with a value of 0.02.

The role of the $z$ component of the magnetic dipole can be
determined by looking at the difference in scattering between the
positive and negative $x$ directions. For example the Opposite
configuration at 1540 nm has a relative scattering difference of
14\% along these two directions. This difference is induced by
interference between a $y$-component electric dipole $d_{y}$ and a
$z$-component magnetic dipole $m_{z}$. Consequently, we can infer
that this structure possesses a significant $z$-component magnetic
dipole relative to the electric dipole. Again this is confirmed in
Table \ref{table1} where the ratio of magnetic to electric dipole
strength $\beta_{z}/\alpha$ has a value of 0.23.

\begin{table}
\caption{Normalized magnitudes of electric and magnetic dipoles}
\begin{ruledtabular}
\begin{tabular}{lccccccr}
Structure & Wavelength[nm] &$\alpha [s_{1}, s_{2}]$ & $\beta_{x} [s_{1}, s_{2}]$&$\beta_{z} [s_{1}, s_{2}]$\\
\hline Single  & 1264 &1 & $0.25$ &$<0.001$\\
       Square & 1264 & 0.23 &0 &0.001\\
       Parallel & 1440 & 0.45 [1.65, -0.66] &0.08 [1.93, -0.93] & 0.08 [0.68, 0.32]\\
       Parallel & 1208 & 1.02 [0.54, 0.46]  &0.26 [0.56, 0.44]  & 0.01 [4.3, -3.3]\\
       Opposite & 1540 & 0.31 [2.06, -1.07] &0.18 [0.58, 0.42]  &0.07 [0.63, 0.37]\\
       Opposite & 1176 & 0.99 [0.53, 0.47]  &0.02 [9.1, -8.1]   &0.008 [4.9, -3.9]\\
\end{tabular}
\end{ruledtabular}
\label{table1}
\end{table}

Higher order multipoles play a more profound role in the linear
scattering of coupled SRRs since their effective thickness of 80 nm
is much larger than 25 nm of a single SRR. At the shorter wavelength
peaks both coupled configurations possess an electric dipole
comparable to that of the single SRR, however, their scattering
along the $x$ direction $[\theta = 0]$ differ by 12\%. Taking into
account that $\beta_{z}$ is negligible for all three structures this
difference cannot be interpreted by interference between the
electric and magnetic dipoles alone. Consequently, contributions
from higher-order multipoles, such as electric quadrupole, must be
included. It should be mentioned that higher-order multipoles are
also found to affect the optical properties of stereometamaterials
\cite{liu}.

To distinguish the relative contribution from each SRR in the
coupled structures we compute their individual dipoles and normalize
them to the total dipole moment. These values are listed in square
brackets in Table \ref{table1} where values with opposite sign
indicate the relevant dipoles are out of phase. For both Opposite
and Parallel configurations the $y$-component electric dipoles
$d_{y}$ are in phase at the shorter wavelength peaks and out of
phase at longer wavelength peaks. This is consistent with the fact
that a coupled system possesses a high-energy symmetric mode as well
as a lower-energy anti-symmetric mode. The inverse of this rule
applies to the $z$-component magnetic dipoles $m_{z}$ in that they
are out of phase at shorter wavelengths and in phase at longer
wavelengths due to the different symmetry of electric and magnetic
fields. Because of the structural symmetry a more complicated phase
relationship exists between the $x$-component magnetic dipoles
$m_{x}$. For example, at the shorter-wavelength resonance they are
in phase for the Parallel configuration while out of phase for the
Opposite one. Table \ref{table1} further implies that although
electric dipole-dipole interaction dominates the inter-SRR coupling,
magnetic dipole-dipole interaction and in particular the
$m_{x}-m_{x}$ interaction must be included to interpret the fact
that the Opposite configuration presents a larger separation between resonance
peaks than its counterpart.

\section{Conclusions}
This paper clarifies the role of magnetic dipoles in single and
double metallic split-ring resonators (SRRs) by numerically
computing their magnitudes and examining their relative
contributions to the scattering cross section. We propose that the
strength of the magnetic dipoles can be determined by measuring the
scattering cross section along particular directions. It is
demonstrated that the magnetic dipole parallel to the incident
magnetic field plays a significant role for Opposite configuration
at its longer wavelength resonance. Moreover, we found that for
isolated pairs of coupled SRRs higher-order multipoles contribute
considerately to the scattering cross section. By considering the
relative phase between the partial dipoles possessed by individual
SRRs, in coupled configurations, we have shown that the electric
dipole-dipole interaction dominates the coupling between adjacent
SRRs and also that the magnetic dipole-dipole interaction is
responsible for the larger resonance separation in the Opposite
configuration.

This work is supported by the Air Force Office of Scientific
Research (AFOSR), under Grant No. FA9550-07-1-0010 and
FA9550-04-1-0213. J. V. Moloney acknowledges support from the
Alexander von Humboldt foundation.

\section{Appendix}
Using a recently developed classical theory \cite{yong2}, we
numerically studied second-harmonic radiation from an individual SRR
as schematically shown in Figure (1a). The fundamental plane wave
propagates along the positive $x$ direction and has a wavelength of
1264 nm, corresponding to the fundamental resonances excited by a
$y$-polarized light. Its linear scattering cross section in the $yz$
plane is shown in Figure (3a). For comparison we also considered a
gold square of size $220\times220$ nm. Its linear response is quite
close to that of the SRR with a $z$-polarized excitation, that is,
they both possess an almost identical-strength electric dipole and a
negligible magnetic dipole. The strongest electric dipole appears in
the SRR with a $y$-polarized excitation.

Figure (3b) plots second-harmonic radiation in the $yz$ plane.
Because of structural symmetry the gold square presents a dominant
electric quadrupole pattern. On the other hand, magnetic dipoles are
found to contribute considerably to the SRR with the $y$-polarized
excitation, and the radiation exhibits a slightly distorted dipole
pattern. It is mainly due to the fact that the second-harmonic
wavelength is around 600 nm while the lateral size of the SRR is
roughly 200 nm. In addition, although the square has almost
identical linear response as the SRR with $z$-polarized excitation,
its second-harmonic strength is much weaker. This is because its
radiation is dominated by an electric quadrupole, while the SRR
radiation is dominated by the electric and magnetic dipole.

\begin{figure}
\epsfxsize=320pt \epsfbox{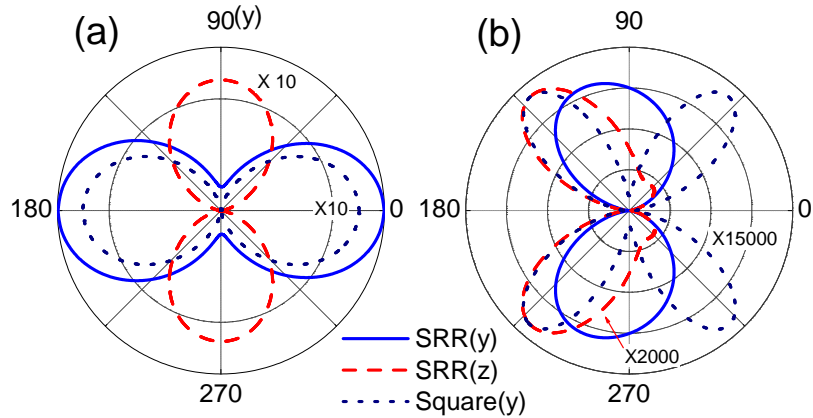} \vspace*{-8.5cm} \caption{Linear
(a) and second-harmonic (b) scattering of the individual split-ring
resonator, specified in Figure (1a), in the $yz$ plane. Both
$y$-polarized (90 degree) and $z$-polarized (zero degree) incidences
are considered. As a reference, a metallic square is also studied.}
\end{figure}

\end{document}